\documentclass[a4paper,twoside]{article}

\usepackage{epsfig}
\usepackage{subcaption}
\usepackage{calc}
\usepackage{amssymb}
\usepackage{amstext}
\usepackage{amsmath}
\usepackage{amsthm}
\usepackage{booktabs}
\usepackage{multicol}
\usepackage{multirow}
\usepackage{pslatex}
\usepackage{apalike}
\usepackage{algorithm2e}
\usepackage{todonotes}
\usepackage{url}
\usepackage[bottom]{footmisc}
\usepackage{SCITEPRESS}     

\begin{document}

\title{Flow Exporter Impact on Intelligent Intrusion Detection Systems}

\author{\authorname{Daniela Pinto\sup{1}\orcidAuthor{0009-0000-3003-6694}, João Vitorino\sup{1}\orcidAuthor{0000-0002-4968-3653}, Eva Maia\sup{1}\orcidAuthor{0000-0002-8075-531X}, Ivone Amorim\sup{2}\orcidAuthor{0000-0001-6102-6165} and Isabel Praça\sup{1}\orcidAuthor{0000-0002-2519-9859}}
\affiliation{\sup{1}Research Group on Intelligent Engineering and Computing for Advanced Innovation and Development (GECAD), School of Engineering, Polytechnic of Porto (ISEP/IPP), 4249-015 Porto, Portugal}
\affiliation{\sup{2}Porto Research, Technology \& Innovation Center (PORTIC), Polytechnic of Porto (IPP), 4200-374 Porto, Portugal}
\email{\{dapsp, jpmvo, egm\}@isep.ipp.pt, ivone.amorim@portic.ipp.pt, icp@isep.ipp.pt}
}

\keywords{Network Intrusion Detection, Network Flow, Dataset, Feature Selection, Machine Learning}

\abstract{High-quality datasets are critical for training machine learning models, as inconsistencies in feature generation can hinder the accuracy and reliability of threat detection. For this reason, ensuring the quality of the data in network intrusion detection datasets is important. A key component of this is using reliable tools to generate the flows and features present in the datasets. This paper investigates the impact of flow exporters on the performance and reliability of machine learning models for intrusion detection. Using HERA, a tool designed to export flows and extract features, the raw network packets of two widely used datasets, UNSW-NB15 and CIC-IDS2017, were processed from PCAP files to generate new versions of these datasets. These were compared to the original ones in terms of their influence on the performance of several models, including Random Forest, XGBoost, LightGBM, and Explainable Boosting Machine. The results obtained were significant. Models trained on the HERA version of the datasets consistently outperformed those trained on the original dataset, showing improvements in accuracy and indicating a better generalisation. This highlighted the importance of flow generation in the model’s ability to differentiate between benign and malicious traffic.}

\onecolumn \maketitle \normalsize \setcounter{footnote}{0} \vfill

\setlength{\tabcolsep}{9pt}
\renewcommand{\arraystretch}{1.15}

\section{\uppercase{Introduction}}
\label{sec:introduction}

Network security threats are an ever-growing concern for all organizations that integrate and interconnect their information systems. For this reason, protecting computer networks has become increasingly important and, in response, Network Intrusion Detection Systems (NIDS) were developed. These systems can detect cyber-attacks by scanning network packets. They identify patterns or anomalies indicating possible security breaches which these systems can then block~\cite{macia-fernandez_ugr16_2018}.

In this context, Artificial Intelligence (AI) has become vital for NIDS, enhancing its ability to adapt to emerging and evolving threats~\cite{THAKKAR2020636}. Historical data is crucial for this process, as it allows Machine Learning (ML) models to learn from past intrusions and predict future attacks with greater accuracy. However, the effectiveness of these ML models hinges on the quality of the datasets used for training. High-quality Network Intrusion Detection (NID) datasets are crucial for ensuring accurate detection and prediction of security threats, as they can aid ML in identifying subtle variations in network traffic, increasing the accuracy of the systems and reducing false positives~\cite{inbook}. 

The datasets developed by the Canadian Institute for Cybersecurity (CIC) are a well-known contribution to this field. Despite their popularity, recent studies have uncovered several limitations and inconsistencies in these datasets, which can negatively impact the performance of ML models~\cite{liu_error_2022}. Besides datasets presenting problems, another challenge lies in the variability of features and flows generated depending on the tools used to process raw network capture packets. The differences in packet aggregation, flow generation, and feature extraction can lead to variations in how the network activity is represented, which in turn can have a significant impact on ML models' generalisation and reliability. However, analysing the impact that flow exporters have on flow generation and ultimately on models' performance has not been extensively explored. Most research on NID datasets has focused either on the process of dataset creation, such as evaluating whether the traffic generated is realistic~\cite{gharib_evaluation_2016}, or on optimising the feature extraction process to generate the dataset~\cite{lanvin_errors_2023}. Consequently, limited attention has been given to how different flow exporters might be impacting the generation of flows and the effects on the dataset quality and performance of the ML models.

This paper aims to study this issue by conducting a comparative analysis of ML models' performance and reliability on datasets derived from the same PCAP files but generated using different flow exporters. By examining the impact of these tools on the resulting datasets, this study highlights how the choice of flow exporter can significantly influence model performance. 

The rest of the paper is structured as follows. Section 2 summarises recent contributions from the research community. Section 3 details the methodology used regarding the selected tool to perform flow extraction and feature exportation. Section 4 elaborates on the obtained results and provides a discussion. Section 5 concludes the paper and presents future work to be performed.

\section{\uppercase{Related Work}}

There has been a growing focus on developing and analysing NID datasets. Several studies have explored the limitations and challenges associated with these datasets. A notable example is the work of Kenyon et al., where the authors evaluated and ranked widely used public NID datasets based on various factors, such as their composition, methodology, maintenance, data quality, and relevance~\cite{kenyon_are_2020}. Their analysis revealed several common issues in publicly available datasets, including difficulties locating reliable datasets, poor methodology in dataset creation, and unclear representation of benign and malicious traffic. Additionally, they highlighted issues with data quality, such as duplicated or unrepresentative samples. The authors also emphasised that many datasets suffer from over-summarisation, which results in the loss of features that limit the ability to scrutinise the original data. Furthermore, a lack of maintenance, particularly in datasets that never receive updates, diminishes their relevance when taking into consideration the constant emergence of new attacks. Another critical finding was that most datasets are limited to specific domains, such as academic networks, and often fail to represent real-world commercial or industrial environments. The authors concluded by proposing best practices for designing NID datasets, stressing the importance of realistic threat coverage and proper anomaly detection techniques.

Conversely, some studies have focused on analysing specific datasets and uncovering issues in their creation. This is especially true for older datasets like KDDCUP'99~\cite{stolfo_cost-based_2000}, where identified shortcomings led to the development of an improved version, NSL-KDD~\cite{tavallaee_detailed_2009}. More recently, the CIC-IDS2017 dataset has also come under scrutiny. Authors such as Engelen et al. have highlighted issues related to the tool CICFlowMeter and the dataset itself, including errors in attack simulations and problems with the feature extraction process~\cite{engelen_troubleshooting_2021}. 

Other studies have focused on evaluating the features used in NID datasets, intending to standardise a feature set. Such an example is the work by Sarhan et al., which proposed a new standardised feature set after evaluating four widely used NID datasets: UNSW-NB15, Bot-IoT, ToN-IoT, and CSE-CIC-IDS2018~\cite{sarhan_towards_2022}. The authors noted that the current NID datasets have unique sets of features making it difficult to compare ML results across datasets and network scenarios. The authors propose two sets of features, one smaller with only 12 and another with 43. They converted the four datasets into new versions, one with 12 features and another with 43 and compared the performance of the new versions with the originals. They noted that the 43 feature datasets consistently outperformed the original datasets in accuracy for binary and multi-class classification. Sarhan et al. conclude that a standardised feature set would enable rigorous evaluation and comparison of ML-based NIDS using different datasets.

A different study on using a different flow exporter tool was produced by Rodríguez et al. where they analysed various ML techniques to determine the most efficient approach for classification, measuring both performance and execution times~\cite{rodriguez_evaluation_2022}. In addition to using the original CIC-IDS2017 dataset, the authors tested using the PCAP files used to generate this dataset on the tool BRO-IDS/Zeek, generating a dataset with 14 features. Their results were favourable, achieving an F1-score above 0.997 with low execution times due to the reduced number of features. They concluded that Zeek allowed them to obtain a better classification than the original dataset features.

\section{\uppercase{Methodology}}

This section describes HERA, the tool used to aid in the creation of labelled NID datasets, the selected flow-based datasets and the considered ML classification models. The analysis was carried out on a common machine with 16 gigabytes of random-access memory and a 6-core central processing unit. The implementation relied on the Python programming language and the following libraries: \textit{numpy} and \textit{pandas} for general data manipulation, and \textit{scikit-learn}, \textit{xgboost}, \textit{lightgbm}, and \textit{interpret} for the implementation of the ML models.

\subsection{\uppercase{HERA}}

Holistic nEtwork featuRes Aggregator (HERA)~\cite{pinto_hera_2024} is a tool that allows users to create flow-based datasets with or without labelling easily. 

In the context of dataset creation, multiple types of tools are used to construct the final dataset. HERA's approach to dataset creation relies on three primary types of tools: \emph{Capturers}, \emph{Flow Exporters}, and \emph{Feature Extractors}. \emph{Capturers} are responsible for recording communication sessions between devices on a network and saving these interactions into files, typically in the PCAP format. Tools such as Wireshark\footnote{\url{https://www.wireshark.org/}} and tcpdump\footnote{\url{https://www.tcpdump.org/}} are common examples of \emph{Capturers}, enabling researchers to preserve packet exchanges for later analysis. \emph{Flow Exporters} aggregate captured network packets into flows, which can be generated directly from network interfaces or PCAP files. These tools group packets according to properties defined by protocols such as NetFlow, jFlow, sFlow, or IPFIX. This type of tool allows for a more efficient, high-level analysis of data, by organizing it into flows. Another important tool category is \emph{Feature Extractors}, which convert raw data into features suitable for analysis. While many \emph{Flow Exporters} also have feature extraction capabilities, it is not required to create flows to be a \emph{Feature Extractor}, only the extraction of features is necessary to have this classification. An example of tools that execute both functions are CICFlowMeter\footnote{\url{https://www.unb.ca/cic/research/applications.html}} and Argus\footnote{\url{https://openargus.org/}}~\cite{pinto_hera_2024}.

In the case of HERA, it utilises PCAP files, that have previously been captured, as input and processes them with the underlying \emph{Flow Exporter} Argus, generating flows and features. For this reason, HERA uses the function of a \emph{Capturer} indirectly by using PCAP files, and has directly the function of a \emph{Flow Exporter} and \emph{Feature Extractor} by using the Argus tool.

HERA takes user input to define parameters when generating the flows. These include arguments that add extra information to the flows, such as packet size and jitter. Additionally, HERA enables the user to choose the interval, in seconds, between flows. This interval specifies the frequency in seconds at which a flow will be generated, provided there is still ongoing activity within the flow during that time. A smaller interval, such as the minimum of 1 second, allows for more frequent updates and provides a more granular view of flow activity, particularly useful in cases where fewer flows might be generated when using the default 60 seconds. Conversely, using larger intervals can reduce the number of flows, which is advantageous when handling larger packet volumes, as expected in scenarios like Distributed Denial-of-Service (DDoS) attacks. Larger intervals help summarize flows with continuous activity by aggregating them into longer-duration flows. Moreover, the user can select which features to be generated, being stored in a CSV file. Also, this tool allows the labelling by providing a CSV file with the ground truth information of the attack traffic. This information includes the start and finish times of the attack, the protocol used, source and destination Internet Protocol (IP) addresses and ports and the label to be applied. 

For the study in question, HERA was used with the PCAP files of the selected datasets, generating the HERA versions of these datasets. Furthermore, by using documentation and ground-truth provided by the selected dataset’s authors, it was possible to label these new versions. 
This permitted to observe the impact that using a different tool has on the generated flows and the prediction of attacks.


\subsection{\uppercase{Datasets}}

The datasets selected for this study were the flow-based NID datasets UNSW-NB15 and CIC-IDS2017. These are datasets with widespread use by the research community with benign and malicious network traffic~\cite{raju_development_2021,wu_intrusion_2022}.
Furthermore, the dataset authors share the PCAP files captured to generate the datasets alongside documentation that allows to establish a ground-truth file that can be used in HERA to generate a new labelled version of these datasets. 

\textbf{UNSW-NB15} was created by the University of New South Wales (UNSW)\footnote{\url{https://www.unsw.edu.au/}}~\cite{moustafa_unsw-nb15_2015} in 2015 with the objective to improve the publicly available datasets, namely KDDCUP’99~\cite{stolfo_cost-based_2000} and NSL-KDD~\cite{tavallaee_detailed_2009} which had been in use since 2000 and 2009 respectively. A criticism provided by the authors is the limited number of attacks that are represented in these datasets and as such in UNSW-NB15, the authors provide 9 attack types to improve diversity. These types are Fuzzers, Analysis, which include port scans, spam and html file penetrations, Backdoors, Denial-of-Service (DoS), Exploits, Generic, Reconnaissance, Shellcode and Worms. The data collected for this dataset was captured in packets and stored in PCAP files by tcpdump. The generated data was simulated by IXIA PerfectStorm\footnote{\url{https://www.keysight.com/us/en/products/network-test/network-test-hardware/perfectstorm.html}}, a tool that emulates real-world network behaviour, in 31 hours on two separate days. For the flow information and features, the PCAP files were used by Argus and BRO-IDS/Zeek\footnote{\url{https://zeek.org/}}, obtaining 49 features with 2 representing attack labels. For the evaluation of this dataset, only a subset of the relevant files of the second day was used in HERA to generate a labelled version.

Concerning the features that were exported, since HERA utilises the same underlying flow exporter, Argus, as one of the employed exporters by the researchers who created the dataset, the features generated for this dataset were easily matched to the UNSW-NB15 dataset. To note, the features that were not exported in comparison to the original dataset, were the ones produced by BRO-IDS/Zeek. So, from the 49 features present in UNSW-NB15, 33 were matched and two were added exclusively by HERA (``FlowID'' and ``Rank''). The matched features are source and destination IP address and port number (``SrcAddr'', ``Sport'', ``DstAddr'' and ``Dport''), the transaction protocol (``Proto''), the transaction state (``State''), the record total duration (``Dur''), the source to destination, and vice-versa, transaction bytes (``SrcBytes'' and ``DstBytes''), time to live values (``sTtl'' and ``dTtl''), packets retransmitted or dropped (``SrcLoss'' and ``DstLoss''), bits per second (``SrcLoad'' and ``DstLoad''), packet count (``SrcPkts'' and ``DstPkts''), Transmission Control Protocol (TCP) window advertisement (``SrcWin'' and ``DstWin''), TCP base sequence number (``SrcTCPBase'' and ``DstTCPBase''), mean of the flow packet size transmitted (``sMeanPktSz'' and ``dMeanPktSz''), jitter (``SrcJitter'' and ``DstJitter''), interpacket arrival time (``SIntPkt'' and ``DIntPkt''), start and last time of the flow (``StartTime'' and ``LastTime''), the TCP connection setup times, in specific the sum of the SYN+ACK and ACK (``TcpRtt''), the time between the SYN and SYN+ACK (``SynAck''), the time between the SYN+ACK and the ACK packets (``AckDat''), and the label containing the attack type, generated from the ground truth (``GTLabel''). Table~\ref{tab_features_unswnb15} provides an overview of the selected features for the ML models' performance evaluation.

\begin{table}[ht]
    \centering
    \caption{Utilised features for UNSW-NB15 dataset.}
    \label{tab_features_unswnb15}
    \begin{tabular}{cc}
        \hline
         \textbf{Original} & \textbf{HERA}  \\
         \textbf{Features} & \textbf{Features}  \\
         \hline
         sinpkt & SIntPkt \\
         dinpkt & DIntPkt \\
         smean & sMeanPktSz \\
         dmean & dMeanPktSz \\
         spkts & SrcPkts \\
         dpkts & DstPkts \\
         sload & SrcLoad \\
         dload & DstLoad \\
         sbytes & SrcBytes \\
         dbytes & DstBytes \\
         sttl & sTtl \\
         dttl & dTtl \\
         \hline
    \end{tabular}
\end{table}

\textbf{CIC-IDS2017} was produced by CIC\footnote{\url{https://www.unb.ca/cic/}}~\cite{sharafaldin_toward_2018} in 2017. Similarly, to the previous dataset, CIC-IDS2017 development was motivated by the need to improve the datasets available to the research community. For this reason, they focused on providing a reliable and up-to-date dataset. This dataset contains attack traffic of DDoS/DoS, Heartbleed, Brute Force, Web Attacks, Infiltrations, Botnets and Port Scans. For the data collected, the authors simulated interaction between 25 users in a controlled environment for 5 days using their tool CICFlowMeter to collect the traffic, transform it into flows and extracting the 84 flow features with 1 representing the attack label. 

Since 2018, CIC-IDS2017 has been scrutinised, having several issues been found in it. A main aspect that has been criticised regarding the dataset stems from problems with the tool used, CICFlowMeter. For instance, the tool has been found to terminate the TCP connections and construct the flow incorrectly due to a timing-related flaw~\cite{engelen_troubleshooting_2021}. Furthermore, labelling errors were encountered in the tool which resulted in incorrect features, feature duplication, feature miscalculation and wrong protocol detection. Other than problems caused by the tool, the dataset has been noted for having high-class imbalance and errors in the attack simulation, labelling and benchmarking~\cite{rosay_cic-ids2017_2021,rosay_network_2022,panigrahi_detailed_2018}.

For the analysis of the dataset, the PCAP files from Monday, Tuesday and Wednesday were used to generate a labelled version of these days using the HERA tool. Since a different tool was used to aggregate the network packets and generate the flow features, Table~\ref{tab_features_cicids2017} provides an overview of the equivalent features that were used to train the ML models. 

\begin{table}[ht]
    \centering
    \caption{Utilised features for CIC-IDS2017 dataset.}
    \label{tab_features_cicids2017}
    \begin{tabular}{cc}
        \hline
         \textbf{Original} & \textbf{HERA}  \\
         \textbf{Features} & \textbf{Features}  \\
         \hline
         fwd\_iat\_mean & SIntPkt \\
         fwd\_iat\_max & SIntPktMax \\
         fwd\_iat\_min & SIntPktMin \\
         bwd\_iat\_mean & DIntPkt \\
         bwd\_iat\_max & DIntPktMax \\
         bwd\_iat\_min & DIntPktMin \\
         active\_mean & Mean \\
         active\_std & StdDev \\
         active\_max & Max \\
         active\_min & Min \\
         \hline
    \end{tabular}
\end{table}

\subsection{\uppercase{Models}}

Four types of ML models were selected for the performance evaluation, based on decision tree ensembles. For each dataset version, distinct models were trained and fine-tuned through a grid search of well-established hyperparameters for multi-class cyber-attack classification with network traffic flows.

To determine the optimal configuration for each model, a 5-fold cross-validation was performed. Therefore, in each iteration, a model was trained with 4/5 of a training set and validated with the remaining 1/5. The selected validation metric to be maximised was the macro-averaged F1-Score, which is well-suited for imbalanced training data~\cite{VitorinoMulticlass}. After being fine-tuned, each model was retrained with a complete training set and evaluated with the holdout set.

\textbf{Random Forest} (RF) is a supervised ensemble of decision trees, which are decision support tools that use a tree-like structure. Each individual tree performs a prediction according to a specific feature subset, and the most voted class is chosen. It is based on the wisdom of the crowd, the concept that the collective decisions of multiple classifiers will be better than the decisions of just one.

The default Gini Impurity criterion was used to measure the quality of the possible node splits, and the maximum number of features selected to build a tree was the square root of the total number of features. Since the maximum depth of a tree was optimized for each training set, its value ranged from 8 to 16 in the different datasets.
Table~\ref{tab_model_rf} summarizes the fine-tuned hyperparameters.

\begin{table}[ht]
\centering
\caption{Summary of RF configuration.}
\label{tab_model_rf}
\begin{tabular}{cc}
\hline
\textbf{Hyperparameter} & \textbf{Value} \\
\hline
Criterion & Gini impurity \\
No. of estimators & 100 \\
Max. depth of a tree & 8 to 16 \\
Min. samples in a leaf & 1 to 4 \\
\hline
\end{tabular}
\end{table}

\textbf{Extreme Gradient Boosting} (XGB) performs gradient boosting using a supervised ensemble of decision trees. A level-wise growth strategy is employed to split nodes level by level, seeking to minimise a loss function during its training.

The Cross-Entropy loss was used with the Histogram method, which computes fast histogram-based approximations to choose the best splits. The key parameter of this model is the learning rate, which controls how quickly the model adapts its weights to the training data. It was optimised to relatively small values for each training set, ranging from 0.05 to 0.25.
Table~\ref{tab_model_xgb} summarizes the hyperparameters.

\begin{table}[ht]
\centering
\caption{Summary of XGB configuration.}
\label{tab_model_xgb}
\begin{tabular}{cc}
\hline
\textbf{Hyperparameter} & \textbf{Value} \\
\hline
Method & Histogram \\
Loss function & Cross-entropy \\
No. of estimators & 100 \\
Learning rate & 0.05 to 0.30 \\
Max. depth of a tree & 4 to 16 \\
Min. loss reduction & 0.01 \\
Feature subsample & 0.8 to 0.9 \\
\hline
\end{tabular}
\end{table}

\textbf{Light Gradient Boosting Machine} (LGBM) also utilises a supervised ensemble of decision trees to perform gradient boosting. Unlike XGB, a leaf-wise strategy is employed, following a best-first approach. Hence, the leaf with the maximum loss reduction is directly split in any level.

The key advantage of this model is its ability to use Gradient-based One-Side Sampling (GOSS) to build the decision trees, which is computationally lighter than previous methods and therefore provides a faster training process. The Cross-Entropy loss was also used, and the maximum number of leaves was optimised to values from 7 to 15 with a learning rate of 0.1 to 0.2. To avoid fast convergences to suboptimal solutions, the learning rate was kept at small values for the distinct datasets.
Table~\ref{tab_model_lgbm} summarizes the hyperparameters.

\begin{table}[ht]
\centering
\caption{Summary of LGBM configuration.}
\label{tab_model_lgbm}
\begin{tabular}{cc}
\hline
\textbf{Hyperparameter} & \textbf{Value} \\
\hline
Method & GOSS \\
Loss function & Cross-entropy \\
No. of estimators & 100 \\
Learning rate & 0.05 to 0.20 \\
Max. leaves in a tree & 7 to 15 \\
Min. loss reduction & 0.01 \\
Min. samples in a leaf & 2 to 4 \\
Feature subsample & 0.8 to 0.9 \\
\hline
\end{tabular}
\end{table}

\textbf{Explainable Boosting Machine} (EBM) is a generalised additive model that performs cyclic gradient boosting with a supervised ensemble of shallow decision trees. Unlike the other three black-box models, EBM is a glass-box model that performs explainable and interpretable predictions.

Despite being a computationally heavy model during the training phase, the maximum number of leaves was limited to 7, and the features used to train this model contribute to a prediction in an additive manner that enables their relevance to be measured and explained quickly during the inference phase.
Table~\ref{tab_model_ebm} summarizes the hyperparameters.

\begin{table}[ht]
\centering
\caption{Summary of EBM configuration.}
\label{tab_model_ebm}
\begin{tabular}{cc}
\hline
\textbf{Hyperparameter} & \textbf{Value} \\
\hline
Loss function & Cross-entropy \\
No. of estimators & 100 \\
Learning rate & 0.10 to 0.15 \\
Max. number of bins & 256 \\
Max. leaves in a tree & 3 to 7 \\
Min. samples in a leaf & 1 to 2 \\
\hline
\end{tabular}
\end{table}

\section{\uppercase{Results and Discussion}}

This section presents the obtained results for the UNSW-NB15 and CIC-IDS2017 datasets, analysing and comparing the ML models' performance on the original network traffic flows with their performance on the flows generated with the HERA tool.

\subsection{\uppercase{UNSW-NB15}}

When analysing the flows obtained with HERA, it was noted that for the same period, the original has, in general, more flows in each classification of traffic as indicated in Table~\ref{tab_flows_unswnb15}. In this particular case, the additional flows are being added by BRO-IDS/Zeek making it possible to note that this flow exporter adds significantly more malicious traffic to the dataset. The biggest discrepancy observed is regarding the traffic identified as ``Generic'', referring to a malicious technique that works against block cyphers, where the original dataset presents 10 times the amount of the HERA version. To note, UNSW-NB15 was produced, in part, with the same underlying flow exporter used by HERA, demonstrating that even adding a different tool leads to different results in flow quantities, even if the same PCAP files were used. 

\begin{table}[ht]
    \centering
    \caption{Flow proportions for UNSW-NB15 dataset.}
    \label{tab_flows_unswnb15}
    \begin{tabular}{ccc}
        \hline
        \textbf{Class} & \textbf{HERA} & \textbf{Original} \\
        \textbf{Name} & \textbf{Flows} & \textbf{Flows} \\
        \hline
        Benign & 726.153 & 893.726 \\
        Exploits & 16.157 & 28.013 \\
        Fuzzers & 12.060 & 14.527 \\
        Generic & 11.468 & 180.076 \\
        Reconnaissance & 7.366 & 9.112 \\
        DoS & 2.294 & 10.549 \\
        Shellcode & 953 & 964 \\
        Analysis & 309 & 1.543 \\
        Backdoor & 232 & 1.425 \\
        Worms & 104 & 110 \\
        \hline
        \textbf{Total} & 777.096 & 1.140.045 \\
        \hline
    \end{tabular}
\end{table}

Due to the imbalanced class proportions of 10 classes of the UNSW-NB15 dataset, the ML models' performance evaluation included macro-averaged metrics to give all classes the same relevance. Several standard evaluation metrics for multi-class cyber-attack classification were computed: accuracy, precision, recall, and F1-Score. Table \ref{tab_results_unswnb15} summarises the obtained results.

\begin{table*}[ht]
\centering
\caption{Obtained results for UNSW-NB15 dataset.}
\label{tab_results_unswnb15}
\begin{tabular}{cccccc}
\hline
\textbf{Dataset} & \textbf{ML} & \textbf{Classification} & \textbf{Macro-averaged} & \textbf{Macro-averaged} & \textbf{Macro-averaged} \\
\textbf{Version} & \textbf{Model} & \textbf{Accuracy} & \textbf{Precision} & \textbf{Recall} & \textbf{F1-Score} \\
\hline
\multirow{4}{4em}{\hfil Original \\ \hfil Flows} & RF & 77.91 & 52.85 & 51.70 & 49.43 \\
& XGB & 78.29 & 59.07 & 51.69 & 49.90 \\
& LGBM & 71.48 & 47.04 & 51.85 & 48.19 \\
& EBM & 77.69 & 55.12 & 50.07 & 48.49 \\
\hline
\multirow{4}{4em}{\hfil HERA \\ \hfil Flows} & RF & 97.68 & 67.81 & 57.07 & 60.50 \\
& XGB & 97.67 & 67.04 & 54.82 & 58.48 \\
& LGBM & 94.88 & 54.91 & 59.13 & 55.76 \\
& EBM & 97.41 & 69.86 & 50.62 & 54.60 \\
\hline
\end{tabular}
\end{table*}

The ML models trained with the original flows obtained reasonably good results in the four metrics. The accuracy of RF, XGB, and EBM was approximately 78\%, with LGBM not being far behind at 71\%, which follows the benchmarks of the current literature~\cite{KILINCER2022107869}. Despite all models reaching high F1-Scores on the majority classes, their macro-averaged scores were below 50\%, which highlights the difficulty in training ML models to distinguish between multiple minority classes of malicious network traffic.

Even though their accuracy was below 80\% on the original version, the models trained with the HERA version achieved between approximately 95\% and 98\% accuracy. Both the precision and recall of all four models were substantially improved, which led to an increase between 7\% and 11\% in their F1-Scores. The best average precision was observed in EBM, 69.86\%, but the best recall and F1-Score were reached by the more lightweight LGBM and RF models, 59.13\% and 60.50\%.

Since the HERA version contained equivalent features that were extracted from the same network packets, better results were achieved by just processing the original data with a tool that uses a single flow exporter as opposed to mixing the flows generated from two different tools. Therefore, this suggests that the approach used to aggregate and convert network packets into flows has a significant impact on ML reliability for NID.

\subsection{\uppercase{CIC-IDS2017}}

For CIC-IDS2017, a similar result to UNSW-NB15 was verified with fewer flows in the dataset as shown in Table~\ref{tab_flows_cicids2017}. Regarding the variation of malicious flows, the DoS Hulk attack presented the most variation with around 70.000 additional flows. In the case of CIC-IDS2017, the flow exporter CICFlowMeter was used. This flow exporter besides having been criticised regarding how it forms the flows, has a different method for limiting flow generation. For this reason, the original dataset has flows that last for over 100 seconds, while the dataset using HERA limits the flow intervals to 60 seconds for this version of the dataset. To note, HERA allows for different flow intervals, but for this study, 60 seconds were chosen as it is the commonly used value.

\begin{table}[ht]
    \centering
    \caption{Flows obtained for CIC-IDS2017 dataset.}
    \label{tab_flows_cicids2017}
    \begin{tabular}{ccc}
        \hline
        \textbf{Class} & \textbf{HERA} & \textbf{Original} \\
        \textbf{Name} & \textbf{Flows} & \textbf{Flows} \\
        \hline
        Benign & 1.186.141 & 1.402.023 \\
        DoS Hulk & 161.225 & 231.073 \\
        DoS Slowhttptest & 8.995 & 5.499 \\
        DoS GoldenEye & 8.972 & 10.293 \\
        DoS Slowloris & 8.729 & 5.796 \\
        FTP Brute Force & 4.003 & 7.938 \\
        SSH Brute Force & 2.959 & 5.897 \\
        Heartbleed & 19 & 11 \\
        \hline
        \textbf{Total} & 1.381.043 & 1.668.530 \\
        \hline
    \end{tabular}
\end{table}

Since the CIC-IDS2017 dataset also contained 8 imbalanced classes, the ML models were also evaluated with the same macro-averaged multi-class classification metrics as before. Table \ref{tab_results_cicids2017} summarises the obtained results.

\begin{table*}[ht]
\centering
\caption{Obtained results for the CIC-IDS2017 dataset.}
\label{tab_results_cicids2017}
\begin{tabular}{cccccc}
\hline
\textbf{Dataset} & \textbf{ML} & \textbf{Classification} & \textbf{Macro-averaged} & \textbf{Macro-averaged} & \textbf{Macro-averaged} \\
\textbf{Version} & \textbf{Model} & \textbf{Accuracy} & \textbf{Precision} & \textbf{Recall} & \textbf{F1-Score} \\
\hline
\multirow{4}{4em}{\hfil Original \\ \hfil Flows} & RF & 96.12 & 97.61 & 76.40 & 84.24 \\
& XGB & 95.95 & 96.43 & 79.65 & 85.70 \\
& LGBM & 92.68 & 89.31 & 79.01 & 82.90 \\
& EBM & 95.44 & 95.30 & 74.92 & 82.64 \\
\hline
\multirow{4}{4em}{\hfil HERA \\ \hfil Flows} & RF & 99.83 & 98.52 & 98.25 & 98.38 \\
& XGB & 99.72 & 96.45 & 96.84 & 96.62 \\
& LGBM & 94.18 & 92.07 & 91.90 & 90.86 \\
& EBM & 99.35 & 95.36 & 91.55 & 93.21 \\
\hline
\end{tabular}
\end{table*}

In contrast with UNSW-NB15, the ML models trained with the original flows of CIC-IDS2017 obtained very high results, due to the more representative classes of this dataset~\cite{VitorinoBenchmark}. The more simple RF model reached the best accuracy and precision, although XGB surpassed it in both recall and F1-Score, indicating a better ability of gradient boosting models to classify multiple cyber-attack classes.

By training with the flows generated by the HERA tool, the models obtained even higher scores. Their accuracy was over 99\% and their F1-Scores over 90\%, which indicates a better generalisation. Despite LGBM and EBM having just a slight improvement in their average precision, all four models reached a substantially better recall, which suggests that the feature extraction performed with the HERA tool led to a better representation of the benign network activity and cyber-attacks present in this dataset. Although the results don't show a drastic improvement, they reinforce the good results obtained by the models associated with CIC-IDS2017, despite the problems identified within the CIC dataset.

The highest overall increase in F1-Score was observed in RF, from 84.24\% to 98.38\%, which corresponds to over 14\%. Therefore, enhancing the conversion of raw data captures into a network flow format enabled the training of more reliable ML models for multi-class cyber-attack classification.

\section{\uppercase{Conclusions}}

This work presented an analysis of the impact that flow exporters have on the generated flow features and on ML models' performance and reliability. The HERA tool was used with the original raw network packets of each selected dataset, creating a new version of these datasets. Multiple models, namely RF, XGB, LGBM, and EBM were trained with the original flows and the HERA flows, and their performance was analysed and compared.

The results obtained demonstrated that RF was the optimal model for both HERA versions of the datasets. While both datasets achieved high accuracy, the F1-Score for the CIC-IDS2017 HERA version was significantly higher, reaching 98.38\% compared to 60.50\% for UNSW-NB15. This indicates that the model performed substantially better on the CIC-IDS2017 dataset, likely due to a higher representation of similar types of attacks. To note, the composition of the datasets might have played a key role in this outcome. UNSW-NB15 contains 6.55\% malicious traffic spread across various attack types, while CIC-IDS2017 includes 14.11\% malicious traffic, with a higher concentration of DoS attacks. Furthermore, the results were also substantially better when the ML models were trained with the HERA flows in comparison to the original, in both datasets. Overall, their macro-averaged F1-Scores were increased between 7\% and 14\%, which indicates a better generalisation. Although the improvement in the ML results for the CIC-IDS2017 dataset was not drastic, these findings demonstrate a new approach to addressing the issues inherent in datasets produced by CIC, particularly those arising from the use of CICFlowMeter. Additionally, by closely aligning the HERA version features with those of the original dataset, it becomes easier to compare results across datasets. Therefore, the results suggest that using an alternative flow exporter, specifically HERA, to aggregate and convert network packets into flows allows ML models to better distinguish between benign network activity and multiple cyber-attack classes.

In the future, it is pertinent to continue analysing the main characteristics of network traffic captures and explore feature engineering techniques to improve the generation of network flows. By providing better datasets with more realistic representations of benign and malicious activity, data scientists and security practitioners will be able to improve the security and trustworthiness of their ML models.

\section*{\uppercase{Acknowledgements}}

This work was supported by the CYDERCO project, which has received funding from the European Cybersecurity Competence Centre under grant agreement 101128052. This work has also received funding from UIDB/00760/2020.

\bibliographystyle{apalike}
{\small
\bibliography{refs}}

\end{document}